\documentclass[pdftex,12pt]{article}
\usepackage{amsmath,amsfonts,amsbsy,latexsym,amssymb,amscd,amstext}
\usepackage{pst-node}
\usepackage{dsfont}
\usepackage{cool}
\usepackage{overpic,youngtab}
\pdfoutput=1
\usepackage{rotating}
\usepackage{tikz}
\usepackage{slashed}

\def\bpm{\begin{pmatrix}}
\def\epm{\end{pmatrix}}
\def\be{\begin{equation}}

\def\ee{\end{equation}}
\def\bea{\begin{eqnarray}}
\def\eea{\end{eqnarray}}
\def\pd{\partial}
\def\a{\alpha}

\def\b{\beta}
\def\G{\Gamma}

\def\g{\gamma}
\def\d{\delta}
\def\m{\mu}

\def\n{\nu}
\def\t{\tau}

\def\l{\lambda}

\def\r{\rho}

\def\s{\sigma}
\def\e{\epsilon}

\def\bma{\begin{pmatrix}}
\def\ema{\end{pmatrix}}

\def\bi{\begin{itemize}}
\def\ei{\end{itemize}}

\begin{document}

		\vspace*{-1cm}
		\phantom{hep-ph/***} 
		{\flushleft
			{{FTUAM-}}
			\hfill{{ IFT-UAM/CSIC-21-84}}}
		\vskip 1.5cm
		\begin{center}
		{\LARGE\bfseries   Some comments on the Hamiltonian for Unimodular Gravity.}\\[3mm]
			\vskip .3cm
		
		\end{center}

		\vskip 0.5  cm
		\begin{center}
			{\large Enrique \'Alvarez and Jes\'us Anero.}
			\\
			\vskip .7cm
			{
				Departamento de F\'isica Te\'orica and Instituto de F\'{\i}sica Te\'orica, 
				IFT-UAM/CSIC,\\
				Universidad Aut\'onoma de Madrid, Cantoblanco, 28049, Madrid, Spain\\
				\vskip .1cm

				\vskip .5cm
				\begin{minipage}[l]{.9\textwidth}
					\begin{center} 
							\textit{E-mail:} 
						\tt{enrique.alvarez@uam.es},
						\tt{jesusanero@gmail.com}
					\end{center}
				\end{minipage}
			}
		\end{center}
	\thispagestyle{empty}
	
\begin{abstract}
	\noindent
	Several alternative formulations of the first order approach to unimodular gravity are presented. There is always a particular one such that it is {\em classically} equivalent to the second order formulation; this we call {\em educated}. It is often at variance with the {\em naive} approach, in which the lagrangian is taken as given exactly by the same expression as in the second order formulation; only the number and character of the  independent variables  changes. Namely, typically some of the momenta are now considered as coordinates. The ensuing Hamiltonians are thereby discussed and their physical differences pointed out.
\end{abstract}

\newpage
\tableofcontents
	\thispagestyle{empty}
\flushbottom

\newpage
 \section{Introduction}
It is well known that every second order equation of motion (EoM) can be written in first order language (FO) just by introducing new dependent  variables. This is usually believed to hold true when there are infinite variables, like in field theory. Nevertheless recently some curious behavior have been pointed out when the gravitational field is considered. Namely, it would seem that there is a crucial difference between lagrangians linear in curvature (Einstein-Hilbert) and lagrangians involving higher orders in curvature. In the latter case, FO is  {\em not} equivalent to the usual second order approach (SO), and the connection field encapsulates many different spin components \cite{Alvarez:2015}.
 \par
 This is indeed a fact when the same lagrangian that is usually worked out in SO is considered as FO; this just means that the metric and the connection are treated as fully independent fields. This will dub herewith as the {\em naive approach}, or {\em naive FO}. The reason is that is it always possible to build up a slightly more complicated FO lagrangian such that its  EoM are completely equivalent as those obtained in the usual SO approach. This we shall dub {\em educated FO}. We shall give many examples in the body of the paper.
 \par
 A simple example, a naive FO for Einstein-Hilbert lagrangian would be
 \be
 S=\int \sqrt{|g|}\,d^n x\, g^{\m\n}\left(\pd_\l \G^\l_{\m\n}-\pd_\m \G^\l_{\n\l}+\G^\s_{\s\l}\G^\l_{\m\n}-\G^\l_{\m\s}\G^\s_{\n\l}\right)
 \ee
 and a educated FO
 \bea
&&  S=\int \sqrt{|g|}\,d^n x\,\bigg\{ g^{\m\n}\left(\pd_\l \G^\l_{\m\n}-\pd_\m \G^\l_{\n\l}+\G^\s_{\s\l}\G^\l_{\m\n}-\G^\l_{\m\s}\G^\s_{\n\l}\right)+\nonumber\\
&&+ \l_\t^{\m\n}\left({1\over 2}\,g^{\t\s}\left(-\pd_\s g_{\m\n}+\pd_\m g_{\n\s}+\pd_\n g_{\s\m}\right) -\G^\t_{\m\n}\right)\bigg\}
 \eea
 where $\l_\t^{\m\n}$is a Lagrange multiplier.
 \footnote{Throughout  this work we follow the Landau-Lifshitz spacelike conventions, in particular the metric is $\eta_{\m\n}=(+,-,-,-)$ and 
 	$R^\m_{~\n\r\s}=\partial_\r \Gamma^\m_{\n\s}-\ldots$
  we omit the factor $-\frac{1}{2\kappa^2}$ in the Einstein-Hilbert action.}
 \par
  We have also recently studied a modification of General Relativity, Unimodular Gravity (\cite{Alvarez:2016} and references therein) in which the set of admissible metrics is restricted to those with unit determinant. The symmetries of the theory are thereby reduced from the set of all diffeomorphisms, $Diff(M)$ to those that preserve the unimodular condition namely the transverse ones $TDiff(M)$. Those transverse vector fields generate the subgroup of {\em volume preserving diffeomorphisms}. We shall always employ the notation
  \be
  \g_{\m\n}
  \ee
  to denote a metric such that
  \be
  \g\equiv \text{det}\,\g_{\m\n}=-1
  \ee
  Sometimes it is useful to generate an unimodular metric out of an  arbitrary one by means of a Weyl rescaling
  \be
  \g_{\m\n}\equiv g^{-1/n}\,g_{\m\n}
\ee
  \par
  In this paper the expressions  {\em linear} or {\em quadratic} are employed always  as referring  to Riemann's curvature tensor.
 
Our aim in this paper is to elaborate on those ambiguities, introducing educated FO and computing in particular the corresponding hamiltonians. The point is that in any diffeomorphism invariant theory the total hamiltonian is a constraint that must be put equal to zero. It is often the case that at least for asymptotically flat space-times the physical energy is embodied in a boundary term, precisely the sort of thing that changes in every "equivalent" formulation of the physical theory.
\par
 A general observation \cite{AAS} is the following. The second order  variation of any lagrangian depending on the metric and the connection field is symbollically  
 \be
 \d S=\int {\d S\over \d \Gamma}{\d \Gamma\over \d g}+{\d S\over \d g}
 \ee
 whereas the first order one read
 \bea
 &&{\d S\over \d \Gamma}=0\nonumber\\
 &&{\d S\over \d g}=0
 \eea
 This clearly shows that $FO$ implies $SO$; the opposite is untrue.
 \par
 The determination of a hamiltonian for the gravitational field is an old problem \cite{Dirac, ADM, Chen}. As has been already pointed out, the bulk hamiltonian vanishes (again, this is actually a generic property of all diffeomorphism invariant theories), and this is the origin of the constraints to be imposed in any canonical quantization of the gravitational field (confer for example \cite{W} and references therein).
 \par
 We will try to be quite specific on the subtle differences between general relativity and unimodular gravity in this respect.
 \par
 Let us stress again that there is also a boundary term which is quite important because it fully determines the numerical value of the energy associated to asymptotically 
at gravitational fields, (ADM)\cite{ADM}. We also would like to give a detailed computation in this case, which could depend on the precise formulation of the theory. Although
 we will devote an initial section to introduce the problem in an explicit {\em physicist} notation using components, the main part of the paper will be written in a more covariant formalism using frame fields and differential forms. This is almost mandatory once higher order (in curvature) lagrangians are considered, in which case the component notation becomes exceedingly cumbersome.
 \section{Theories linear in curvature. The Einstein-Hilbert lagrangian.}
 \par
\subsection{The Einstein-Hilbert lagrangian.}

 The Einstein-Hilbert lagrangian in FO formalism is
 \bea \mathcal{L}_{EH}&=\sqrt{|g|} g^{\m\n}\left(\partial_\l\Gamma^\l_{\m\n}-\partial_\n\Gamma^\l_{\l\m}+\Gamma^\l_{\t\l}\Gamma^\t_{\m\n}-\Gamma^\l_{\t\m}\Gamma^\t_{\l\n}\right)\eea
\par
 A  related action principle (educated version)  that would be equivalent to Einstein-Hilbert's would read
 \bea \label{EFO}\mathcal{L}_{EH}&=-\Gamma^\l_{\m\n}\partial_\l\left(\sqrt{|g|} g^{\m\n}\right)+\Gamma^\l_{\l\m}\partial_\n\left(\sqrt{|g|} g^{\m\n}\right)+\sqrt{|g|} g^{\m\n}\left(\Gamma^\l_{\t\l}\Gamma^\t_{\m\n}-\Gamma^\l_{\t\m}\Gamma^\t_{\l\n}\right)\eea
 The variation respect to the metric reads
 \bea&& -\Gamma^\l_{\m\n}\partial_\l\left(\sqrt{|g|}\left( \frac{1}{2}g^{\m\n}h-h^{\m\n}\right)\right)+\Gamma^\l_{\l\m}\partial_\n\left(\sqrt{|g|}\left( \frac{1}{2}g^{\m\n}h-h^{\m\n}\right)\right)-\nonumber\\
 && -\sqrt{|g|}\left(\left(\Gamma^\l_{\t\l}\Gamma^\t_{\m\n}-\Gamma^\l_{\t\m}\Gamma^\t_{\l\n}\right)-\frac{1}{2}g_{\m\n}g^{\a\b}\left(\Gamma^\l_{\t\l}\Gamma^\t_{\a\b}-\Gamma^\l_{\t\a}\Gamma^\t_{\l\b}\right) \right)h^{\m\n}=0\eea
 under integration by parts we recover the Einstein field equation
 \bea &\frac{1}{2} R h-R_{\m\n}h^{\m\n}=-G_{\m\n}h^{\m\n}=0\nonumber\\
 \eea
 and the variation respect to the connection of the lagrangian reads
 \bea&&\Big[-\partial_a(\sqrt{|g|}g^{bc})+\d^b_a\partial_d(\sqrt{|g|}g^{cd})-\nonumber\\
 &&-\frac{1}{2}\sqrt{|g|}g^{\m\n}\Big(\d_\m^c\Gamma^b_{a\n}+\d_\m^b\Gamma^c_{a\n}+\d_\n^b\Gamma^c_{a\m}+\d_\n^c\Gamma^b_{a\m}-\d_a^c\Gamma^b_{\m\n}-\d_a^b\Gamma^c_{\m\n}-\d^b_\m\d_\n^c\Gamma^\l_{a\l}-\d^c_\m\d_\n^b\Gamma^\l_{a\l}\Big)\Big]A^a_{bc}=0\nonumber\\\eea
 which using
 \be\partial_\l\left(\sqrt{|g|} g^{\m\n}\right)=\sqrt{|g|}\left(g^{\m\n}\Gamma^\t_{\t\l}-g^{\t\n}\Gamma^\m_{\t\l}-g^{\m\t}\Gamma^\n_{\t\l}\right)\ee
 reduces to zero.
This fact shows \cite{AASG} that the linear lagrangian does not need an educated form, in the sense that naive FO is already equivalent to SO. 
 \par
 Let us examine now what is the situation in the unimodular setting. Consider a linear unimodular metric in FO, $\g_{\m\n}$ and some associated torsionless connection, $\omega^\m_{\n\l}$. We are interested in the action
 \be
 S\equiv \int d^n x\, \g^{\n\s}\left(\pd_\m \omega^\m_{\n\s}-\pd_\s \omega^\m_{\n\m}+\omega^\m_{\l\m}\omega^\l_{\n\s}-\omega^\m_{\l\s}\omega^\l_{\n\m}\right)
 \ee
 where the covariant derivative acts on the covariant indices only. Please note that the Levi-Civita connection associated to the unimodular metric satisfies
 \be
 \omega_\s\equiv \omega^\l_{\l\s}=0
 \ee
 and
 \be
 \omega^\s\equiv \g^{\a\b} \omega^\s_{\a\b}=\pd_\l \omega^{\s\l}
 \ee

 \par
 In terms of an arbitrary non-unitary metric \cite{Alvarez:2016}, $g_{\m\n}$
 \be
 \g_{\m\n}\equiv g^{-{1\over n}}g_{\m\n}
 \ee
 this formulation introduces a redundant Weyl gauge symmetry
 \be
 g_{\m\n}\rightarrow \Omega^2(x)\,g_{\m\n}
 \ee
 
 The Levi-Civita connection associated to $\g_{\m\n}$ is
 \be
 \omega^\m_{\n\r}=\Gamma^\m_{\n\r}+{1\over 2n}\,g^{\m\l}\left({\pd_\l g\over g} g_{\n\r}-{\pd_\n g\over g} g_{\l\r}-{\pd_\r g\over g} g_{\n\l}\right)
 \ee
 where $\Gamma^\m_{\n\r}$ is the Levi-Civita connection associated to the general metric $g_{\m\n}$; that is, Christoffel's symbols.
 
 \par
 Next, we present our notation for the ADM formalism \cite{ADM, W, Poisson}. The metric tensor $g_{\m\n}$ of spacetime induces a metric $h_{\m\n}$ on the spatial hypersurface $\sum_t$,
 \be h_{\m\n}=g_{\m\n}+n_\m n_\n\ee
 where $n_\m$ is the unit normal. The extrinsic curvature tensor of the spatial hypersurface $\sum_t$ is defined as
 \be K_{\m\n}=\nabla_\m n_\n+n_\m a_\n=\nabla_\m n_\n+n_\m n^\l\nabla_\l n_\n\ee
 the extrinsic curvature can be written as the Lie derivative of the induced metric $h_{\m\n}$
 on $\sum_t$ along the unit normal $n$ to $\sum_t$
 \be\label{K} K_{\m\n}=\frac{1}{2}\mathcal{L}_n h_{\m\n}\ee
 and the decomposition of the scalar curvature R of spacetime can be written as
 \be R=^{(3)}R+K_{\m\n}K^{\m\n}-K^2+2\nabla_\m(n^\m K-a^\m)\ee
 
 In the given ADM coordinate base, the components of the metric of spacetime read
 \be g_{00}=N^2\quad g_{0i}=N_i\quad g_{ij}=h_{ij}\ee
 the extrinsic curvature tensor \eqref{K} is written as
 \be  K_{ij}=\frac{1}{2}\mathcal{L}_n h_{ij}=\frac{1}{2N}\left(\partial_t h_{ij}-D_i N_j-D_j N_i\right)\ee
 \begin{enumerate}
 \item {\em Einstein Hilbert in second order.}
 \par
 What happens in second order UG \cite{Unruh} is that the lapse is not an independent dynamical variable, because
 \be
 N^2 |h|=1
 \ee
 where 
 \be
 h\equiv \det\,g^{(n-1)}_{ij}\equiv \det\,h_{ij}
 \ee
 This means that it is not compulsory to impose the hamiltonian constraint
 \be
 {\cal H}=0 
 \ee
  \bea
 &\{{\cal H}(x),{\cal H}(x^\prime)\}=\left({\cal H}^i(x)+{\cal H}^i(x^\prime)\right)\pd_i\d(x-x^\prime)\nonumber\\
&\{{\cal H}_i(x),{\cal H}(x^\prime)\}\sim{\cal H}(x)\pd_i \d(x-x^\prime)\nonumber\\
&\{{\cal H}_i(x),{\cal H}_j(x^\prime)\}\sim{\cal H}_i(x^\prime)\pd_j \d(x-x^\prime)+{\cal H}_j(x)\pd_i\d(x-x^\prime)
 \eea
but only the weaker condition
  \be
  {\cal H}=  \l
  \ee
  where $\l$ is deternined by the physical boundary conditions. This is the usual unimodular setting \cite{Alvarez:2015} in hamiltonian language.
 \item{\em Einstein Hilbert in naive first order.}
\par 
 It has already been pointed out that it has been proved in \cite{AASG} that even the naive FO Einstein-Hilbert is equivalent to the usual SO Einstein's equations. As for the hamiltonian, there are several possibilities. We could, for example start with the lagrangian of GR in naive FO ADM form  (and neglecting boundary terms) 
\be \mathcal{L}_{EH}= N\sqrt{h}\,\Big[{}^{(3)}R+K_{ij}K^{ij}-K^2\Big]\ee
where the variables are the spatial metric $h_{ij}$ and the extrinsic curvature, $K_{ij}$. With these assumptions all momenta vanish and the hamiltonian just coincides with the potential.
 \item{\em Einstein Hilbert in educated first order.}
\par
Let us instead start with the lagrangian of GR in educated FO ADM form \cite{W, Poisson} (and neglecting boundary terms) 
\be \label{Se}\mathcal{L}_{EH}= N\sqrt{h}\,\Big[{}^{(3)}R+\frac{1}{N}K^{ij}\left(\partial_t h_{ij}-D_i N_j-D_j N_i\right)-K_{ij}K^{ij}-2KT+K^2\Big]\ee
where $D_i$ is the induced covariant derivative  in the 3-manifold $\Sigma_t$, and we have defined $T\equiv \frac{1}{2N}h^{ij}\left(\partial_t h_{ij}-D_i N_j-D_j N_i\right)$. The EoM for the field $K^{ij}$ implies
\be \frac{1}{N}\left(\partial_t h_{ij}-D_i N_j-D_j N_i\right)-2K_{ij}-2h_{ij}T+2h_{ij}K=0\ee
then 
\be K_{ij}=\frac{1}{2N}\left(\partial_t h_{ij}-D_i N_j-D_j N_i\right)\ee
with $T=K$, note if we reintroduce the expresion of $K_{ij}$ in \eqref{Se}, we recover the standard SO lagrangian in ADM variables, the conjugate momenta $p^{ij}$ are

\bea
N_i &&\rightarrow p^i=0\nonumber\\
K_{ij}&&\rightarrow p^{ij}=0\nonumber\\
h_{ij}&&\rightarrow p^{ij}=\sqrt{h}K_{ij}\eea
therefore the Hamiltonian 
\be H_{EH}=-N\sqrt{h}{}^{(3)}R+p^{ij}\left(D_i N_j+D_j N_i\right)+\frac{N}{\sqrt{h}}\left(p_{ij}p^{ij}+p^2\right)\ee

 \item{\em Unimodular Einstein Hilbert in second order.}
\par
Again here the unimodular constraint 
\be
N\sqrt{h}=1
\ee
implies that the unimodular lagrangian  reads
\be \mathcal{L}_{UG}= {}^{(3)}R+\frac{1}{N}K^{ij}\left(\partial_t h_{ij}-D_i N_j-D_j N_i\right)-K_{ij}K^{ij}-2KT+K^2\ee
and the conjugate momenta $p^{ij}$ are
\bea
N_i &&\rightarrow p^i=0\nonumber\\
K_{ij}&&\rightarrow p^{ij}=0\nonumber\\
h_{ij}&&\rightarrow p^{ij}=\frac{1}{N}K_{ij}\eea
and the Hamiltonian
\be H_{UG}={}^{(3)}R+p^{ij}\left(D_i N_j+D_j N_i\right)+N^2\left(p_{ij}p^{ij}+p^2\right)\ee
\item{\em Unimodular educated first order.}
\par
The unimodular version of \eqref{EFO} yields
\be \mathcal{L}_{UE}=-\Gamma^\l_{\m\n}\partial_\l\left( \g^{\m\n}\right)+\Gamma^\l_{\l\m}\partial_\n\left( \g^{\m\n}\right)-\g^{\m\n}\left(\Gamma^\l_{\t\m}\Gamma^\t_{\l\n}-\Gamma^\l_{\t\l}\Gamma^\t_{\m\n}\right)\ee
and the conjugate momenta $p^{ij}$ are
\bea
\Gamma^\l_{\m\n} &&\rightarrow p_\l^{\m\n}=0\nonumber\\
\g_{00}&&\rightarrow p^{00}=-\Gamma^0_{00}+\Gamma_{\l 0}^\l\nonumber\\
\g_{0i}&&\rightarrow p^{0i}=-\Gamma^0_{0i}+\Gamma_{\l i}^\l\nonumber\\
\g_{ij}&&\rightarrow p^{ij}=-\Gamma^0_{ij}\eea
and the Hamiltonian reads
\be H_{UE}=\Gamma^i_{\m\n}\partial_i \g^{\m\n}-\Gamma^\l_{\l\m}\partial_ i \g^{i\m}+\g^{\m\n}\left(\Gamma^\l_{\t\m}\Gamma^\t_{\l\n}-\Gamma^\l_{\t\l}\Gamma^\t_{\m\n}\right)\ee

\end{enumerate}
\subsection{ Schr\"odinger's lagrangian.}
The Einstein-Hilbert lagrangian, can be written as
\bea \mathcal{L}_{EH}&=\sqrt{|g|}R=\partial_\l\left(\sqrt{|g|} g^{\m\n}\Gamma^\l_{\m\n}\right)-\partial_\n\left(\sqrt{|g|} g^{\m\n}\Gamma^\l_{\l\m}\right)+\mathcal{L}_{S}\eea
where 
\be
\partial_\l\left(\sqrt{|g|} g^{\m\n}\right)=\sqrt{|g|}\left(g^{\m\n}\Gamma^\t_{\t\l}-g^{\t\n}\Gamma^\m_{\t\l}-g^{\m\t}\Gamma^\n_{\t\l}\right)
\ee
then, up to a total derivative
\bea \mathcal{L}_{S}&=\sqrt{|g|}g^{\m\n}\mathfrak{L}_{\m\n}=\sqrt{|g|}g^{\m\n}\left(\Gamma^\l_{\t\m}\Gamma^\t_{\l\n}-\Gamma^\l_{\t\l}\Gamma^\t_{\m\n}\right)\eea
which is just the $\G\G$ Schr\"odinger's \cite{Schrodinger} lagrangian. It is then plain that the Einstein-Hilbert and Schr\"odinger's lagrangian differ by a total derivative; so that they yield the same equations of motion when considered in second order. 
\par
The energy-momentum tensor of Schr\"odinger's lagrangian is 
\bea T^{\m}_\n&&=\frac{\partial \mathcal{L}_{S}}{\partial (\partial_\m g_{\a\b})}\partial_\n g_{\a\b}-\mathcal{L}_{S}\d^{\m}_\n=\nonumber\\
&&=\frac{\sqrt{g}}{2}\Big[2\G_{\a\b}^\m-\G^\m_{\r\s}g^{\r\s}g_{\a\b}-\d^\m_\a\G^\l_{\b\l}-\d^\m_\b\G^\l_{\a\l}\Big]g^{\a\b}_{~~,\n}-\d^{\m}_{\n}\sqrt{|g|}g^{\a\b}\left(\Gamma^\l_{\t\a}\Gamma^\t_{\l\b}-\Gamma^\l_{\t\l}\Gamma^\t_{\a\b}\right)\nonumber\\\eea
It is curious that it corresponds to the so called Einstein energy pseudo tensor  \cite{Chen}, which  in first order formalism,  reduces to
\bea T^{\m}_\n&&=-\mathcal{L}_{S}\d^{\m}_\n\eea
\par
What about Schr\"odinger's lagrangian considered as a first order one? The dependence on the variables $\Gamma^\a_{\b\g}$ and $g_{\m\n}$ is algebraic, so that the lagrangian is equivalent to the hamiltonian
\be
H\equiv V=-L
\ee
 \bi
\item  Let us now  include physical  sources for the graviton, $T^{\m\n}$ and for the connection field, $ j^{\b\g}_\a$ in Schr\"odinger's lagrangian
\be\label{Ls}
\mathcal{L}_{Smatter}\equiv \sqrt{g}\left(~g^{\m\n}\mathfrak{L}_{\m\n}+g^{\m\n} T_{\m\n}+\Gamma^\a_{\b\g}j^{\b\g}_\a\right)
\ee
the variation with respect to the metric \footnote{ There is a small subtlety here. The variation of the scalar $T\equiv g^{\m\n} T_{\m\n}=g_{\m\n} T^{\m\n}$ is
	\be
	\d g^{\m\n} T_{\m\n}\neq \d g_{\m\n} T^{\m\n}
	\ee
	The explanation is that it is {\em not} equivalent to assume $\d T^{\m\n}=0$ than to assume $\d T_{\m\n}=0$. Indeed
	\be
	\d T^{\m\n}=\d g_{\a\l} T^\l_\b+\d g_{\b\l} T^\l_\a+g_{\a\m}g_{\b\n} \d T^{\m\n}
	\ee
	Here we are assuming $\d T_{\m\n}=0$.
}
of \eqref{Ls} yields
\be
\d \mathcal{L}_S=\sqrt{|g|}\left(-\mathfrak{L}^{\m\n}+{1\over 2} \mathfrak{L} g^{\m\n}-T^{\m\n}+{1\over 2 } T g^{\m\n}+{1\over 2} g^{\m\n} \G^\a_{\b\g} j^{\b\g}_\a\right)\d g_{\m\n}
\ee
the trace of the EoM yields
\be
(n-2)(\mathfrak{L}+T)  +n\G^\a_{\b\g} j^{\b\g}_\a=0
\ee
in the absence of sources this implies 
\be
\mathfrak{L}=0
\ee
neglecting for the time being the connection source, the EoM for the graviton field reads
\be
\mathfrak{L}_{\m\n}+T_{\m\n}=0
\ee


The variation of \eqref{Ls} with respect to the connection yields
\be
\d \mathcal{L}_S=\Bigg\{\frac{1}{2}g^{\m\n}\Big[\d_\m^c\Gamma^b_{a\n}+\d_\m^b\Gamma^c_{a\n}+\d_\n^b\Gamma^c_{a\m}+\d_\n^c\Gamma^b_{a\m}-\d_a^c\Gamma^b_{\m\n}-\d_a^b\Gamma^c_{\m\n}-\d^b_\m\d_\n^c\Gamma^\l_{a\l}-\d^c_\m\d_\n^b\Gamma^\l_{a\l}\Big]+j^{bc}_a\Bigg\}A^a_{bc}
\ee
where $A^a_{bc}=\d\G^a_{bc}$, then
\be \label{2}2\Gamma^{b|c}_{a}+2\Gamma^{c|b}_{a}-\d_a^c g^{\m\n}\Gamma^b_{\m\n}-\d_a^b g^{\m\n}\Gamma^c_{\m\n}-2g^{bc}\Gamma^\l_{a\l}+2j^{bc}_a=0\ee
trace \eqref{2}, with $\d_c^a$
\be g^{\m\n}\left[(1-n)\Gamma^c_{\m\n}+2j^{c\l}_{\l}\right]=0\ee
Now we trace \eqref{2}, with $g_{cb}$
\be (4-2n)\Gamma^{\l}_{a\l}-2g_{ac} g^{\m\n}\Gamma^c_{\m\n}+2g_{bc}j^{bc}_a=0\ee
\par
\item
Let us work out the linear approximation
\bea
&&g_{\m\n}=\eta_{\m\n}+\kappa h_{\m\n}\nonumber\\
&&\G^\a_{\b\g}=0+A^\a_{\b\g}
\eea
where we just have seen that
\be
\mathfrak{L}^{(L)}=\frac{1}{2}h T-h^{\m\n}T_{\m\n}+\frac{1}{2}h A^\l_{\m\n}j^{\m\n}_\l
\ee
\par
At any rate, it is plain that without sources
\be
\mathfrak{L}^{(L)}_{\a\b}=0
\ee
it does then seem impossible to recover Newton's equation in the appropiate linear limit in FO. We have just seen that in  SO we recover Einstein's equations exactly. The reason for this apparent contradiction is that the difference between Schr\"odinger and Einstein-Hilbert lagrangians is a total derivative only when considered in SO, but it is not when considered in FO.
\par
This then illustrates a dramatic instance of a non-equivalence of FO and SO approaches in a theory of gravity linear in curvature. The standard lore \cite{AASG} was that FO and SO were equivalent for theories linear un curvature (such as the standard Einstein-Hilbert lagrangian), and nonequivalence appears only in theories involving higher powers of Riemann's tensor.

In conclusion, whereas the $\Gamma-\Gamma$ lagrangian correctly reproduce Einstein's equations when considered in SO, it predicts flat space in vacuum as the only solution in FO. No gravitational waves exist in this formulation!
\ei
\section{Theories quadratic in curvature.} 
Quadratic theories can be written in the general form
\be
S=\frac{1}{2}\int d^n\, x \,\sqrt{|g|}\,\bigg\{R^\m\,_{\n\r\s}[\G]\,P_{\m\m^\prime}\,^{\n\n^\prime\r\r^\prime\s\s^\prime}[g]R^{\m^\prime}\,_{\n^\prime\r^\prime\s^\prime}[\G]\bigg\}
\ee
where the tensor $P_{\m\m^\prime}\,^{\n\n^\prime\r\r^\prime\s\s^\prime}[g]$ depends only on the metric $g_{\a\b}$ and its inverse $g^{\a\b}$.
Let us work out the basic example where
\be
P_{\m\m^\prime}\,^{\n\n^\prime\r\r^\prime\s\s^\prime}[g]=\d_\m^{\m^\prime}\d_{\m^\prime}^\n g^{\r\r^\prime} g^{\s\s^\prime}%
\ee
other contractions of Riemann's tensor  can be worked out along  similar rules. An educated  first order version of the action principle is given by promoting $R^\m\,_{\n\r\s}$ to an independent variable together with the connection $\G$ and the metric tensor $g$.
\bea
&&S=\frac{1}{2}\int d^n\,x\,\sqrt{|g|}\,\bigg\{R^\m\,_{\n\r\s}g^{\r\a} g^{\s\b} R^\n\,_{\m\a\b}+2 R_\m\,^{\n\r\s}\left(\pd_\r \G^\m_{\n\s}-\pd_\s\G^\m_{\n\r}+\G^\m_{\l\r}\G^\l_{\n\s}-\G^\m_{\l\s}\G^\l_{\n\r}\right)\bigg\}\nonumber\\
\eea
In FO it is not necessary to introduce auxiliary fields as in \cite{Deruelle}.  We shall assume that the field $R_\m\,^{\n\r\s}$ has the symmetries of Riemann's tensor.

In fact the Lagrangian EoM ensure that it is given on shell by the Riemann tensor corresponding to the dynamical connection $\Gamma$.
\be
R^\m\,_{\n\r\s}[\G]\equiv\pd_\r \G^\m_{\n\s}-\pd_\s\G^\m_{\n\r}+\G^\m_{\l\r}\G^\l_{\n\s}-\G^\m_{\l\s}\G^\l_{\n\r}
\ee
\par
The canonical momenta are given by
\bea
g_{\m\n}&&\rightarrow\quad p^{\m\n}\sim 0\nonumber\\
R^\m\,_{\n\r\s}&&\rightarrow\quad p_\m\,^{\n\r\s}\sim 0\nonumber\\
\G^\m_{\n i}&&\rightarrow\quad p^{\n i}_\m=\sqrt{|g|} \left(R_\m\,^{\n 0 i}+R_\m\,^{i 0 \n}\right)
\eea
This object is  symmetric in $(\n i)$ and as usual, greek indices run from $ (0 \ldots  n-1)$ and latin indices from the middle of the alphabet run from $1\ldots n-1$. When contracted with some other tensor with those symmetries it is not necessary to make those explicit.Let us define the auxiliary variable
\be
q_\m^{\n i}\equiv \sqrt{|g|} \left(R_\m\,^{\n 0 i}- R_\m\,^{i 0 \n}\right)
\ee
in conclusion
\be
R_\m\,^{\n 0 i}={1\over 2\sqrt{|g|}}\left(q_\m^{\n i}- p_\m^{\n i}\right)
\ee
now we need to substitute certain components of Riemann's tensor by the corresponding momenta. The computation becomes heavy and we refrain from reproducing it here; we shall give a simplified treatment using differential forms momentarily.

\section{Covariant approach in terms of the frame field and the spin connection.}
When discussing a hamiltonian formalism, it is unavoidable to introduce a non-covariant distinction between space and time. This can be done however in such a way that as many symmetries as possible as are respected. The time direction will be characterized by a vector  field
\be Z\equiv Z^\m\partial_\m\ee
then acting on any exact form
\be
i_Z d\a=\pounds_Z \a-d i_Z \a\equiv \dot{\a}-d i_Z \a
\ee
($\pounds_Z\a$ is the generalization of the concept of {\em time derivative}). It is always possible to locally choose an adapted coordinates, such that, $Z=\partial_t$  with $i_Z dt=1$. It is then natural \cite{Chen}, to define  the time and space projections on an arbitrary  form $\a$  as
\bea
\hat{\a}&&\equiv i_Z \a\nonumber\\
\underline{\a}&&\equiv \a-dt\wedge \hat{\a}_t
\eea

The induced projections on the exterior differential read

\be
d\a\equiv \underline{d\a}+dt\wedge \widehat{d\a}=\underline{d\a}+dt\wedge \left(\dot{\a}-d\hat{\a}\right)
\ee
it will also prove convenient to decompose the differential operator $d=dx^\m\wedge\partial_\m=dt\wedge\partial_t+dx^k\wedge\partial_k$, and we define
\bea
d\equiv dt\wedge \hat{d}+\underline{d}
\eea
with $\hat{d}\equiv \pounds_Z$
\par
After this small introduction, let us write down the variation of the first order lagrangian. Our purpose is to get an expression for the associated hamiltonian by particularizing later for an explicit form of the variation.
\bea
&&\d \mathcal{L}_{FO}=\sum\Big[d\left(\d \phi^k\wedge p_k\right)+\d\phi^k\wedge\frac{\d\mathcal{L}_{FO}}{\d\phi^k}+\frac{\d\mathcal{L}_{FO}}{\d p_k}\wedge \d p_k\Big]
\eea
this formula does not assume anything about the variation. Consider the particular case of a time translation; that is a Lie derivative along the vector field $Z$
\be
\d\phi^k=\pounds_Z\phi^k
\ee
it follows in general that 
\bea
d\, i_Z \,\mathcal{L}_{FO}&&=\pounds_Z\mathcal{L}_{FO}=\sum\Big[d\left(\pounds_Z \phi^k\wedge p_k\right)+\pounds_Z\phi^k\wedge\frac{\d\mathcal{L}_{FO}}{\d\phi^k}+\frac{\d\mathcal{L}_{FO}}{\d p_k}\wedge \pounds_Z p_k\Big]
\eea
this means that there is a first integral
\be
H(Z)\equiv \sum \pounds_Z\phi^k\wedge p_k -i_Z\, \mathcal{L}_{FO}
\ee
satisfies the identity
\be\label{dH}
-d H(Z)=\sum \Big[\pounds_Z\phi^k\wedge\frac{\d\mathcal{L}_{FO}}{\d\phi^k}+\frac{\d\mathcal{L}_{FO}}{\d p_k}\wedge \pounds_Z p_k\Big]
\ee
it is a conserved current on shell, which explicit expression is
\bea\label{HZ}
&&H(Z)=\sum \Big[d(i_Z \phi^k\wedge p_k)+i_Z\phi^k\wedge dp_k+d\phi^k\wedge i_Zp_k+i_Z \Lambda\Big]
\eea
can be written like a displacement vector plus a total differential
\be
H(Z)\equiv Z^\m H_\m +d B(Z)
\ee
where 
\be\label{B} B(Z)=\sum i_Z \phi^k\wedge p_k\ee
compare the differential of this expression $d H= dZ^\m \wedge H_\m+ Z^\m dH_\m$ with \eqref{dH}, we learn that
\bea\label{H}
&&Z^\m H_\m = \sum \Big[-i_Z\phi^k\wedge\frac{\d\mathcal{L}_{FO}}{\d\phi^k}+\frac{\d\mathcal{L}_{FO}}{\d p_k}\wedge i_Z p_k\Big]
\eea
so that $H_\m$ itself vanishes on shell, and all contribution to the energy comes from the boundary term, $B(Z)$.
\subsection{\em Einstein-Hilbert theory.}
\par
In terms of the frame one-forms \footnote{Flat or Lorentz indices are raised or lowered with the flat metric $\eta_{ab}$; whereas Einstein or curved indices do that with the metric $g_{\m\n}$.
}
\be
e^a\equiv e^a_\m dx^\m
\ee
and the curvature two-form
\be
R_{ab}\equiv {1\over 2} R_{ab\m\n} dx^\m\wedge dx^\n
\ee
the Einstein-Hilbert action can be written \cite{Alvarez:2015} as the integral over spacetime 
\be
S= \int e^a\wedge e^b\wedge *R_{ab}=\int R_{ab}\wedge *\left( e^a\wedge e^b\right)
\ee
where the curvature is expressed in terms on the connection one-forms
\be
\omega_{ab}\equiv \omega_{ab\m} dx^\m
\ee
note $\omega_{ab\m}=-\omega_{ba\m}$, then
\be
R_{ab}\equiv d\omega_{ab}+ \omega_{ac}\wedge \omega^c\,_b
\ee
In second order formalism the connection one-forms are determined by the torsionless condition
\be\label{T}
d e^a+\omega^a\,_b\wedge e^b=0
\ee
but in this paper we would like to stick to the first order formalism, in which $\omega$ is an independent field.
\par
Let us explain in detail how this comes about.
\bea
S&&={1\over 2} \int  \e_{abcd}\, e^a\wedge e^b\wedge R^{cd}={1\over 4} \int\,\e_{abcd}e^a_\m e^b_\n R^{cd}\,_{\r\s} dx^\m\wedge dx^\n\wedge dx^\r\wedge dx^\s\nonumber\\\eea
but
\be
dx^\m\wedge dx^\n\wedge dx^\r\wedge dx^\s=d^nx\sqrt{|g|} \e^{\m\n\r\s}
\ee
\bea
S&&=  {1\over 4}\int d^nx\,\sqrt{|g|} \e^{\m\n\r\s}\e_{\m\n cd}R^{cd}\,_{\r\s}= {1\over 2}\int d^nx\,\sqrt{|g|} \d^{\r\s}_{ cd}R^{cd}\,_{\r\s}=\int d^nx\sqrt{|g|}  R\nonumber\\
\eea

The conjugate momenta are given by
\bea\label{p}
e^a &&\rightarrow p_a\equiv {\pd \mathcal{L}\over \pd de^a}=0\nonumber\\
\omega_{ab}&&\rightarrow p^{ab}\equiv{\pd \mathcal{L}\over \pd d \omega_{ab}}=*\left(e^a\wedge e^b\right)
\eea
then the Legendre transform is performed though the construct \cite{Chen}
\be
\Lambda\equiv \sum d\phi^i\wedge p_i-\mathcal{L}=d\omega_{ab}\wedge p^{ab}-\left(d\omega_{ab}+\omega_a\,^e\wedge \omega_{eb}\right)\wedge *\left(e^a\wedge e^b\right)=-  \omega_a\,^e\wedge \omega_{eb}\wedge p^{ab}
\ee
this defines a first order lagrangian, namely
\bea
&\mathcal{L}_{FO}\equiv \sum d\phi^i\wedge p_i-\Lambda=d\omega^{ab}\wedge p_{ab}+\omega^{a e}\wedge\omega_e\,^b\wedge p_{ab}=R^{ab}\wedge p_{ab}
\eea
which EoM read
\bea
&&{\d \mathcal{L}_{FO}\over \d e^a}=0\nonumber\\
&&{\d \mathcal{L}_{FO}\over \d \omega^{ab}}=-d p_{ab}+ \omega^{c}\,_{a}\wedge p_{cb}-\omega^{c}\,_{b}\wedge  p_{ac}\nonumber\\
&&{\d \mathcal{L}_{FO}\over \d p_{ab}}=R^{ab}
\eea

In our case
\be
H(Z)\equiv \sum \pounds(Z) \phi^i\wedge p_i -i_Z\, L_{FO}=\pounds(Z)\omega^{ab}\wedge p_{ab}-i_Z\left(p_{ab}\wedge R^{ab}\right)
\ee
the displacement piece of the hamiltonian is then given in this language by the vectorvalued three-form
\be H_\m=-\omega^{ab}_{~~\m}\wedge\left(-d p_{ab}+ \omega^{c}\,_{a}\wedge p_{cb}-\omega^{c}\,_{b}\wedge  p_{ac}\right)+R^{ab}\wedge p_{ab\m}\ee
it follows that the one-forms $p_{ab\m}$ are given in terms of the frame by
\be p_{ab\m}=\frac{1}{2}\e_{abcd}\left(e^c_\m e^d-e^d_\m e^c\right)\ee
this then the form that the hamiltonian constraint and momentum constraints take in this formalism. The boundary term, \eqref{B}, is given by the two form
\be
B=i_Z\omega^{ab}\wedge p_{ab}=\frac{1}{2}\e^{ab}\,_{cd}\omega_{ab\m}Z^\m\wedge e^c\wedge e^d
\ee
Particularizing for Schwarzschild's metric
\be ds^2=f^2(r)dt^2-\frac{dr^2}{f^2(r)}-r^2 d\theta^2-r^2\sin^2\theta d\phi^2\ee
where $f(r)=\sqrt{1-\frac{r_s}{r}}$, with $r_s=2GM$, then the frame field read
\bea \label{F}
&&e^0=f(r)dt\nonumber\\
&&e^1=\frac{dr}{f(r)}\nonumber\\
&&e^2=r d\theta\nonumber\\
&&e^3=r\sin\theta d\phi
\eea
in such a way that the nontrivial connection one-forms are given by
\bea\label{O}
&&\omega^0_{~1}=f'(r) e^0\nonumber\\
&&\omega^2_{~1}=\frac{f(r)}{r}e^2\nonumber\\
&&\omega^3_{~1}=\frac{f(r)}{r}e^3\nonumber\\
&&\omega^3_{~2}=\frac{\cot\theta}{r}e^3
\eea
The integral of the boundary over the 2-sphere $S^2_\infty$ at infinity 
\bea
&t=\text{constant}\nonumber\\
& r=R\uparrow\infty
\eea
reads

\be \int_{S^2_\infty}B=r_s\pi dt(Z)=\frac{\kappa^2 M_{\odot}}{4}dt(Z)\label{BEH}\ee
\subsection{\em Unimodular gravity.}
\par
In \cite{Alvarez:2015} it has been proved that if the simplest FO lagrangian 
\be\label{simple}
S_{UG}=  \int R_{ab}\wedge *\left( \hat{e}^a\wedge \hat{e}^b\right)
\ee
is chosen where
\be
\hat{e}^a\equiv e^{- 1/n} e_a
\ee
then there is on shell a non-vanishing torsion, which however vanishes in the Weyl gauge $e=1$. The reason is that this lagrangian is Weyl invariant with inert spin connection, but the torsionless condition is not.
\par
One way out is to impose a nontrivial Weyl transformation of the spin connection in such a way that the torsionless condition is Weyl invariant. The resulting lagrangian has the drawback in that it dependds not only on $\hat{e}_a$, but also on $e$.
\par

The simplest alternative would probably be to consider again the lagrangian \eqref{simple}, but with Weyl dependent connection. This is what will be implicitly done here, although we shall  not be interested in the Weyl symmetry in this paper.
Then
\be
S_{UG}= \int R_{ab}\wedge *\left( e^{-2/n}\,e^a\wedge e^b\right)
\ee
in this case the canjugate momenta is
\be
p^{ab}\equiv *\left( e^{-2/n}\,e^a\wedge e^b\right)
\ee
therefore the Legendre transform result
\be
\Lambda=d\omega^{ab}\wedge p_{ab}-\left(d\omega^{ab}+\omega^{ac}\wedge\omega_c\,^b\right)\wedge *\left( e^{-2/n}\,e^a\wedge e^b\right)=-\omega^{ac}\wedge\omega_c\,^b\wedge p_{ab}
\ee
so that
\be
{\cal L}_{FO}=d\omega^{ab}\wedge p_{ab}+\omega^{ac}\wedge\omega_c\,^b\wedge p_{ab}
\ee
which is exactly the same as before. The displacement piece of the hamiltonian is then
given in this language by the vector-valued three-form 
\be H_\m=-\omega^{ab}_{~~\m}\wedge\left(-d p_{ab}+ \omega^{c}\,_{a}\wedge p_{cb}-\omega^{c}\,_{b}\wedge  p_{ac}\right)+R^{ab}\wedge p_{ab\m}\ee
it follows that the one-forms $p_{ab\m}$ are given in terms of the frame by
\be p_{ab\m}=\frac{1}{2}\e_{abcd}\left(\hat{e}^c_\m \hat{e}^d-\hat{e}^d_\m \hat{e}^c\right)\ee
this then the form that the hamiltonian constraint and momentum constraints take in this formalism. The explicit expression obviously coincide with the general-relativistic ones in the Weyl gauge e = 1, and  when the vector $Z$ is chosen in the ADM manner like
\be Z=n\,{\pd\over \pd x^0}+N^i\,{\pd \over \pd x^i}
\ee
we recover the unimodular constraint $N\sqrt{h}=1$.

The boundary term, \eqref{B}, in the hamiltonian is now
\be
B_{UG}=i_Z\omega_{ab}\wedge p^{ab}=\frac{1}{2}\e^{ab}\,_{cd}\widehat{\omega}_{ab\m}Z^\m \wedge \hat{e}^c\wedge \hat{e}^d
\ee
particularizing to Schwarzschild's metric
\be e=r^2\sin\theta\ee
using the frame field, \eqref{F}, we obtain the connection
\be \widehat{\omega}_{01\m}Z^\m=\Big[-\frac{1}{2r}f(r)+f'(r)\Big]f(r)dt(Z)\ee
\be
B_{UG}=\frac{1}{2}\frac{1}{r\sin^{1/2}\theta}\Big[-\frac{1}{2r}f(r)+f'(r)\Big]f(r)dt(Z)r^2\sin\theta\, d\theta\wedge d\phi
\ee
but 
\be f'(r)=\frac{1}{2f(r)}\frac{r_s}{r^2}\ee
over the sphere at infinity
\bea
\int_{S^2_\infty}B_{UG}&&=-\frac{1}{4}dt(Z)\int_{S^2_\infty}\sin^{1/2}\theta \, d\theta\wedge d\phi=-\sqrt{2\pi}\Big[\Gamma\left(\frac{3}{4}\right)\Big]^2 dt(Z)
\eea
In is instructive in this context to consider the unimodular frame \cite{S}, due to Schwarzschild \footnote{
Please beware of an annoying arratum in equantion (8) of the proprint version of \cite{S}.}
himself
\be ds^2=f_0dt^2-f_1dx_1^2-f_2\left(\frac{dx_2^2}{1-x_2^2}+(1-x_2^2)dx_3^2\right)\ee
where $x_1=r^3/3$, $x_2=-\cos\theta$, $x_3=\phi$ and
\bea 
f_0(x_1)&&=1-\frac{2GM}{(3x_1+b)^{1/3}}\nonumber\\
f_1(x_1)&&=\frac{(3x_1+b)^{-4/3}}{1-\frac{2GM}{(3x_1+b)^{1/3}}}\nonumber\\
f_2(x_1)&&=(3x_1+b)^{2/3}
\eea
where $b$ is a constant of integration and $f_0f_1f_2^2=1$, then the frame field reads
\bea
e^0&&=\sqrt{f_0}dt\nonumber\\
e^1&&=\sqrt{f_1}dx_1\nonumber\\
e^2&&=\frac{\sqrt{f_2}}{\sqrt{1-x_2^2}}dx_2\nonumber\\
e^3&&=\sqrt{(1-x_2^2)f_2}dx_3\eea

The boundary term, \eqref{B}, in the hamiltonian is now
\be
B_{UGSch}=\frac{1}{2}\e^{ab}\,_{cd}\widehat{\omega}_{ab\m}Z^\m \wedge \hat{e}^c\wedge \hat{e}^d
\ee
we can derive
\be de^0=-\frac{GM}{f_2}dt\wedge e^1\ee
but $de^0+\omega^0_{~1}\wedge e^1=0$, we obtain the connection one-form
\be \widehat{\omega}_{01\m}Z^\m=\frac{GM}{f_2}dt(Z)\ee
and
\be
B_{UGSch}=\frac{GM}{2}dt(Z) dx_2\wedge dx_3
\ee
over the sphere at infinity
\bea
\int_{S^2_\infty}B_{UG}&&=\frac{GM}{2}dt(Z)\int_{S^2_\infty}\sin\theta \, d\theta\wedge d\phi=\pi r_s dt(Z)
\eea
this result reproduces \eqref{BEH}.
\subsection{\em Schr\"odinger's lagrangian.}
\par
The Einstein-Hlbert action can be written
\be S=\int R_{ab}\wedge *\left(e^a\wedge e^b\right)=\frac{1}{2}\int \epsilon^{abcd}\left(d\omega_{ab}+ \omega_{af}\wedge \omega^f\,_b\right)\wedge \left(e_c\wedge e_d\right)\ee
let us write $d\left(e_c\wedge e_d\right)=d e_c\wedge e_d-e_c\wedge d e_d$ and using the torsionless condition, \eqref{T}, obtain
\bea
S&&=\frac{1}{2}\int d^n x\epsilon^{abcd}\Big\{d\Big[\omega_{ab}\wedge \left(e_c\wedge e_d\right)\Big]+2\omega_{ab}\wedge \omega_{cf}\wedge e^f\wedge e_d+ \omega_{af}\wedge \omega^f\,_b\wedge e_c\wedge e_d\Big\}\nonumber\\\eea
in terms of the frame componnets on the connection field
\bea
S&&=\frac{1}{2}\int d^n x\epsilon^{abcd}\Big\{2\omega_{abu}\omega_{cfv} e^u\wedge e^v\wedge e^f\wedge e_d+ \omega_{afu}\omega^f_{~bv} e^u\wedge e^v\wedge e_c\wedge e_d\Big\}\nonumber\\\eea
where we neglecting the total derivative, and 
\be e^u\wedge e^v \wedge e^a\wedge e^b=d^n xe\,\e^{uvab}\ee
we get
\bea
S&&=\frac{1}{2}\int d^n x\, e\epsilon^{abcd}\Big\{2\omega_{ab}^{~~u}\omega_{c}^{~fv}\e_{uvfd}+\omega_{af}^{~~u}\omega^{f\cdot v}_{~b}\e_{uvcd}\Big\}\eea
but
\bea 
\epsilon^{abcd}\e_{uvfd}&=\d^{abc}_{uvf}\nonumber\\
\epsilon^{abcd}\epsilon_{uvcd}&=2\d^{ab}_{uv}
\eea
and finally
\bea
S&&=3\int d^n x\,e\Big[\omega_{ca}^{~~a}\omega_{b}^{~cb}+\omega_{abc}\omega^{cba}\Big]
\eea
this is Schr\"odinger's lagrangian in terms of forms. In FO there are no non-vanihing momenta, so that the hamiltonian is just
\bea
H(Z)&&=-i_Z \mathcal{L}\eea
Thereby there is no  boundary term.
\par
Incidentaly, the value of the Schr\"odinger's lagrangian for
Schwarzschild's solution is
\be \mathcal{L}=\frac{6r-4r_s+2r\cot^2\theta}{r^3}\ee
\subsection{\em Quadratic theories.}
\par
In terms of the one form
\be
\G^\m_\n\equiv \G^\m_{\n\l} dx^\l
\ee
and the two-form
\be
R^\m\,_\n\equiv {1\over 2}R^\m\,_{\n\r\s}dx^\r\wedge dx^\s
\ee
the preceding lagrangian reads
\be
S=\int  R^\m\,_\n\wedge *R^\n\,_\m+2 R^\m\,_\n\wedge * \left( d\G^\n_\m+\G^\n_\r\wedge \G^\r_\m\right)
\ee
and the EoM read
\bea
&&{\d S\over \d R^\m\,_\n}=R^\n\,_\m-\left(d\G^\n_\m+\G^\n_\r\wedge \G^\r_\m\right)=0\nonumber\\
&&{\d S\over \d \G^\m\,_\n}=*2d R^\n\,_\m+R^\n\,_\s\wedge *\G^\s_\m-R^\l\,_\m\wedge *\G^\n_\l =0
\eea
because $d *=* \d$.The corresponding momenta read
\bea
&&p\equiv {\pd L\over \pd d R^\m\,_\n}=0
\nonumber\\
&&p^\n\,_\m\equiv {\pd L\over \pd d \G^\m\,_\n}= 2 * R^\n\,_\m
\eea
therefore the Legendre transform result
\be
\Lambda= d\G^\m\,_\n\wedge p^\n\,_\m-\frac{1}{4}p^\n\,_\m\wedge * p^\m\,_\n-p^\m_\n\wedge\left(d\G^\n_\m+\G^\n_\r\wedge \G^\r_\m\right)=-\frac{1}{4}p^\n\,_\m\wedge * p^\m\,_\n-p^\m_\n\wedge\left(\G^\n_\r\wedge \G^\r_\m\right)
\ee
and the FO lagrangian in this language reads
\be
\mathcal{L}_{FO}=\frac{1}{4}p^\n\,_\m\wedge * p^\m\,_\n+p^\m_\n\wedge\left(d\G^\n_\m+\G^\n_\r\wedge \G^\r_\m\right)
\ee
therefore the  FO EoM read
\bea
&&{\d \mathcal{L}_{FO}\over \d \G^\m\,_\n}=d p^\n\,_\m+p^\n\,_\s\wedge \G^\s_\m-p^\l\,_\m\wedge \G^\n_\l 
\nonumber\\
&&{\pd \mathcal{L}_{FO}\over \pd p^\m\,_\n}=*\frac{1}{4}p^\n\,_\m+\left(d\G^\n_\m+\G^\n_\r\wedge \G^\r_\m\right)
\eea
using the spacetime decomposition, \eqref{H}
\bea
Z^\m H_\m &&= -i_Z \G^\m_\n\wedge\left(d p^\n\,_\m+p^\n\,_\s\wedge \G^\s_\m-p^\l\,_\m\wedge \G^\n_\l \right)+\left(*\frac{1}{4}p^\n\,_\m+\left(d\G^\n_\m+\G^\n_\r\wedge \G^\r_\m\right)\right)\wedge i_Z p^\m_\n\nonumber\\
\eea
then
\bea
H_\l &&= - \G^\m_{\n\l}\wedge\left(d p^\n\,_\m+p^\n\,_\s\wedge \G^\s_\m-p^\l\,_\m\wedge \G^\n_\l \right)+2\left(*\frac{1}{4}p^\n\,_\m+\left(d\G^\n_\m+\G^\n_\r\wedge \G^\r_\m\right)\right)\wedge \e^{\a\b}_{~~\l\s} R^\m_{~\n\a\b} dx^\s
\nonumber\\\eea
because 
\be i_Z p^\m_\n=i_Z\e^{\a\b}_{~~\r\s} R^\m_{~\n\a\b}dx^\r\wedge dx^\s=2\e^{\a\b}_{~~\r\s} R^\m_{~\n\a\b}Z^\r dx^\s\ee
as usual in diffeomorphism invariant theories $H_\l$ itself vanishes on shell, and all contribution to the energy comes from the boundary term, \eqref{B}
\bea
B&&=i_Z\G^\m\,_\n\wedge p^\n\,_\m=2 \G^\m\,_{\n\l} Z^\l\wedge*R^\n\,_\m=\nonumber\\
&&=\frac{1}{2e}\G^\m\,_{\n\l} Z^\l\wedge \e_{~\m}^{\n~\a\b}R_{\a\b\r\s}dx^\r\wedge dx^\s
\eea
now we need integrate over the sphere 
\bea
B&&=\frac{1}{e}\G^\m\,_{\n\l} Z^\l\wedge \e_{~\m}^{\n~\a\b}R_{\a\b 23}dx^2\wedge dx^3
\eea
particularizing to Schwarzschild metric $R_{2323}=\left[1-f^2(r)\right]\sin^2\theta$
then
\bea
B&&=\frac{2}{e}\G^\m\,_{\n\l} Z^\l\wedge \e_{~\m}^{\n~23}\left[1-f^2(r)\right]\sin^2\theta d\theta\wedge d\phi
\eea
again for Schwarzschild
\bea
B&&=\frac{2}{e}\Big[\G^0\,_{10}-\G^1\,_{00}\Big] dt(Z)\left[1-f^2(r)\right]r^2\sin^3\theta d\theta\wedge d\phi
\eea
but $\G^0\,_{10}=\frac{f'(r)}{f(r)}$ and $\G^1\,_{00}=f^4(r)\G^0\,_{10}$
\bea
B&&=\frac{2f'(r)}{f(r)}\Big[1-f^4(r)\Big] dt(Z)\left[1-f^2(r)\right]\sin^2\theta d\theta\wedge d\phi
\eea
with $f(r)=\sqrt{1-\frac{r_s}{r}}$ over the sphere at infynity, with the usual asignment 
\be dt(Z)=1\ee
it yields vanishing hamiltonian energy. As is well-known there are other formulations \cite{Deser} that assign a finite energy to those configurations.

  \section{Conclusions}
We have computed the hamiltonian corresponding to different first order versions of unimodular gravity.
\par
It must be stressed that the {\em naive} approach, in which the lagrangian is taken as given exactly by the same expression as in the second order approach, with the proviso that the role of the independent variables is changed, namely the connection field and the metric field are now to be treated as independent, this naive approach, we stress, is not always equivalent to the more usual second order one.
\par
Theories linear in curvature have been studied, both the one that corresponds to the standard Einstein-Hilbert lagrangian, as well as the one related to  Schr\"odinger's version quadratic in the connection field. This last version is particularly interesting insofar as it can be viewed as giving a rationale for (Einstein's)  energy-momentum pseudotensor. While it is in fact true that both versions differ by a total derivative, it is not less true that one of the most interesting aspects of the hamiltonian in generally covariant framework is precisely the boundary term, also a total derivative, and precisely this boundary term usually depends on those total derivatives.
\par
Theories quadratic in curvature have also been considered. In this case the proliferation of indices quickly becomes overewhelming. At any rate, the usual formalism is unsatisfactory here insofar as it yields vanishing energy in this case.
\par
It has been found often convenient in this paper to use the language of frame fields and differential forms. This is the more true when dealing with  theories with lagrangians quadratic in curvature, although the formalism  saves much space even in simpler contexts. 
\par
We are working in a frame formulation of the ideas in \cite{Deser} in order to give a satisfactory definition of energy in the quadratic case. We hope to be able to report on it in due time.
\section{Acknowledgements}
This work has been partially supported by the Spanish Research Agency (Agencia Estatal de Investigacion) through the PID2019-108892RB-I00/AEI/ 10.13039/501100011033 grant as well as the IFT Centro de Excelencia Severo Ochoa SEV-2016-0597 one, and the European Union's Horizon 2020 research and innovation programme under the Marie Sklodowska-Curie grants agreement No 674896 and No 690575. 

\newpage
\appendix


%
%
\newpage

\end{document}